\documentclass[showpacs,preprintnumbers,nofootinbib, amsmath,amssymb,twocolumn]{revtex4}
\usepackage{amsmath}
\usepackage{graphicx}
\usepackage{aas_macros}
\pdfoutput=1

\begin{document}

\title{Holes in the static Einstein universe and the model of 
the cosmological voids}

\author{Andrzej Odrzywolek}
% \address{}
 \homepage{http://www.ribes.if.uj.edu.pl/}
 \email{odrzywolek@th.if.uj.edu.pl}
\affiliation{
M. Smoluchowski Institute of Physics\\ 
Jagiellonian University\\  
Reymonta 4\\ 
30-059 Krakow\\ 
Poland
}%

\date{\today}

\begin{abstract}
Spherically symmetric, static model of the cosmological voids is constructed
in the framework of the Tolman-Oppenheimer-Volkov equation
with the cosmological constant. Extension of the Tooper result (dimensionless form
 of the TOV equation) is provided
for non-zero $\Lambda$. Then, the equation is simplified
in $\alpha \to 0$, $\lambda \to 0$, $\lambda/\alpha = const$ regime, suitable
for largest structures in $\Lambda$-dominated universe. 
Voids are treated as an underdensity regions   
in the static Einstein universe. 
Both overdensity and underdensity (relative to static universe) solutions exist. 
They are 
identified with standard astrophysical spherical objects and voids, respectively. Model is tested
against observed properties (the radius - the central density relation) and density profiles of voids. 
Analytical formulae for radial density contrast profile and radii of the voids are derived. 
Some consequences for cosmological n-body simulations are suggested. Hints on the
dark matter/dark energy EOS filling the voids are provided.
\end{abstract}


\maketitle

\section{Introduction}

In both Newton's
and Einstein's theories of gravitation a large set of static\footnote{More precisely,
quasi-static, as all these objects evolve.} astrophysical
objects well approximated in spherical symmetry exist: moons, planets, brown dwarfs, stars,
globular clusters, galactic clusters, and so on. Spherical shape of these objects is
the most obvious and the most common effect of the universal
attractive force. However, existence of roughly spherically
symmetric \citep{1987ApJ...314..493K} underdensity regions in the large-scale 
matter distribution in the universe,
named voids \citep{2001ApJ...557..495P}, was an unexpected discovery 
\citep{1981ApJ...248L..57K,1982Natur.300..407Z}. We will show, in the framework of the non-linearly 
perturbed static Einstein universe (thereafter s.E.u.), the existence of static, spherically symmetric objects devoid 
of matter relative to the surrounding. They belong to the same
family of solutions as normal overdensity objects ranging from planets to galactic clusters.
Spherical underdensity regions (''voids'') are however directly related to the existence of cosmological
constant $\Lambda$. Due to very small numerical value of $\Lambda$, they exist only for the largest-scale structures of universe: the voids.
For smaller objects, $\Lambda$ provides negligible contribution. With $\Lambda=0$ only overdensity
static solutions do exist.

Despite extensive literature on the static, spherically symmetric
solutions of Einstein equations (see e.g: \cite{PhysRevD.69.104028,PhysRevD.67.104015,boonserm:044024} 
and references therein) this particular branch was not discussed in the literature
known to author. Papers of \cite{2004ragt.meet...75H}
and \cite{2008PhRvD..77f4008B} deal with similar model, but only the 
overdensity solutions were found and analyzed.

The article goals are: 
\begin{enumerate}
\item{to show the existence of the {\em static} solutions with {\em positive} (!)
pressure gradient
}
\item{to derive and analyze basic equations for the generalized polytropic fluid spheres
with non-zero $\Lambda$; this generalize result of \citet{1965ApJ...142.1541T}
}
\item{to emphasize existence (and explore properties) of the two classes of the solutions; 
static Einstein universe is a boundary between them
}
\item{to show repulsive ''gravitational force'' acting on the test particles in the vicinity of the ''hole''}
\item{to use the above solutions in the spherically symmetric static model of the cosmological
voids}
\item{to test the model against observed density profiles and $\rho_c-D_{void}$ relation
for the voids}
\end{enumerate}

Meanwhile, renewed interest in static Einstein universe 
resulted in a large number of recently published papers, see e.g.
\cite{ebert:064029, bohmer:084005, 2007CQGra..24.6243P, herrera:027502, 2008PhRvD..77l7502L, goswami:044011, 2008PhRvD..78d4028G, 2009PhRvD..79f4009S, 2009PhRvD..79f7504B, 2009arXiv0905.3546L, 2009PhRvD..80d3528C, 2009arXiv0909.2821W}
and references in these papers

Article is organized as follows. In Sect.~\ref{sect:SEU} we remind well known properties of the 
s.E.u,
whose physical properties (density, pressure and radius) are used in next section, Sect.~\ref{sect:TOVlambda}, 
to simplify notation. 
In the Sect.~\ref{sect:TOVlambda} we analyze family of solutions to the Tolman-Oppenheimer-Volkov
(see e.g. \cite{Stuchlik}) equation with the cosmological constant. 
Some surprising properties of the underdensity solutions are presented in Sect.~\ref{sect:HoleProperties}.
Generalization of the \citet{1965ApJ...142.1541T} result (valid only for polytropic EOS) is presented
in Sect.~\ref{sect:Tooper-lambda}.
Finally, in Sect.~\ref{sect:TOV-lambda-fit}, attempt is made to fit solutions of the TOV-$\Lambda$ 
(Tooper-$\lambda$) equation to the observed
density profiles of the voids. Possible consequences of the results are outlined in the last section, 
Sect.~\ref{sect:Conclusions}.

\section{Static Einstein Universe } \label{sect:SEU}

Traditionally, static universe is discussed
in the context of the general Friedmann-Lemaitre-Robertson-Walker
(FLRW) solution, see \cite{lrr-2001-1}. \citet{RevModPhys.5.62} put it into separate
class ''E''.

For a brief introduction into s.E.u we restrict ourselves to the metric representing
3-sphere with radius $R(t)$:
\begin{equation}
\label{3-sphere}
ds^2 = -dt^2 + R(t)^2 \; 
\left[
d \chi^2 + \sin^2{\chi} \, d \Omega^2
\right]
\end{equation}
Here, $d \Omega^2 = \sin^2{\theta} d \phi^2 + d \theta^2$,
and $\chi, \theta, \phi$ are spherical coordinates
on unit 3-sphere.
The Einstein equations are:
\begin{subequations}
\label{closed_univ}
\begin{equation}
\label{closed_univ_1}
3 \left( \frac{\dot{R}}{R} \right)^2 + \frac{3}{R^2} = 8 \pi G \rho + \Lambda
\end{equation}
\begin{equation}
\label{closed_univ_2}
3 \frac{\ddot{R}}{R} + 4 \pi G (\rho+ 3 p)=\Lambda
\end{equation}
\end{subequations}

Additionally, we have to specify properties of matter filling the universe.
We consider Equation Of State (thereafter EOS)
of the barotropic ideal fluid:
\begin{equation}
\label{barotropic_fluid}
p=p(\rho),
\end{equation}
where $p$ - pressure, $\rho$ - energy density.

Some of the examples widely used in cosmology are
dust ($p=0$) and  relativistic gas ($p=\rho/3$). Later in the article we will use
sub-class of barotropic EOS, {\em polytropic gas}:
\begin{equation}
\label{polytropic_fluid}
p= K \rho^{\gamma}, \quad \gamma = 1+ \frac{1}{n}.
\end{equation}

If we assume radius of the 3-sphere constant in \eqref{closed_univ}:
$$
R(t)=R_E
$$
then the Einstein's equations reduce to a simple system of algebraic equations: 
\begin{subequations}
\label{static_system}
\begin{equation}
\label{static_system_1}
8 \pi G \; \rho_E = \frac{3}{{R_E}^2} - \Lambda
\end{equation}
\begin{equation}
\label{static_system_2}
8 \pi G\; p_E = - \frac{1}{{R_E}^2} + \Lambda
\end{equation}
\end{subequations}
where EOS is given by \eqref{polytropic_fluid}.

\begin{figure}
\includegraphics[width=0.5\textwidth]{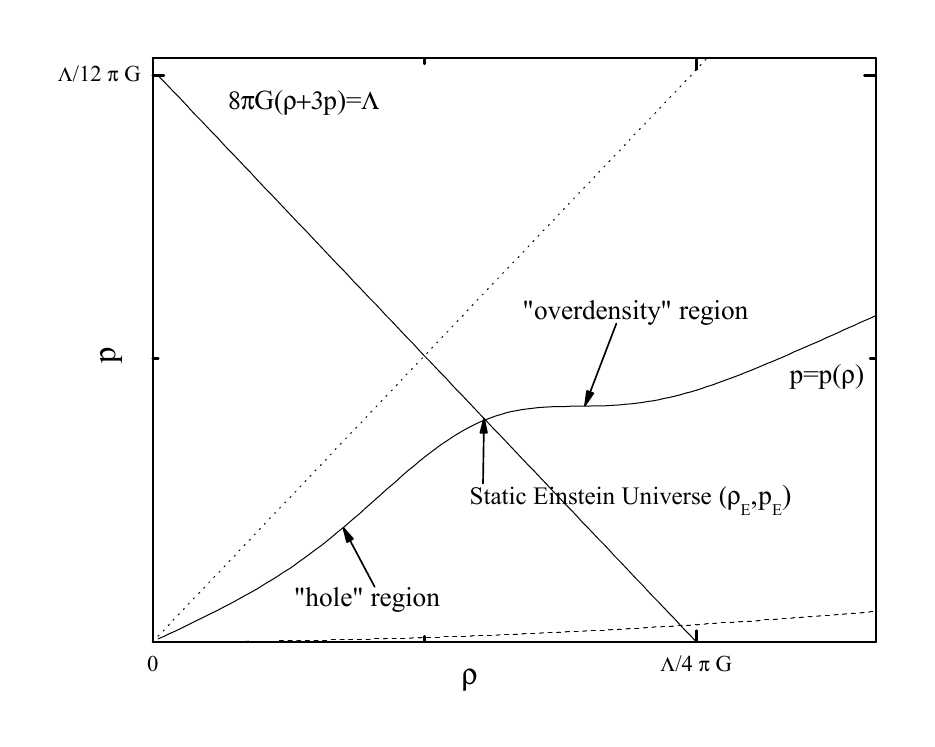}
\caption{\label{static_sys_fig} Graphical solution of  \eqref{static_source} with EOS \eqref{polytropic_fluid}. Dashed line
represents eq.~\eqref{static_source}, and solid line corresponds to EOS. Dash-dotted line
shows relativistic EOS $p=\rho/3$. Static 3-sphere solution (line crossing) exist for any ''normal'' 
EOS.} 
\end{figure}

Let us note, that linear combination of eqs.~\eqref{static_system}
gives (see also \eqref{closed_univ_2}):
\begin{equation}
\label{static_source}
4 \pi G \; (\rho_E+3\, p_E) =  \Lambda.
\end{equation}
The gravitational field ''source'' $\rho+3\, p$
required for static solution does not depend on the EOS. For all 
possible static Einstein universes is exactly the same. We will show later, 
that quantity $\rho_E+3\, p_E$ is very important for inhomogeneous static solutions.

If we specify standard EOS\footnote{With $p,\rho>0$ and $p \leq \rho/3$.}
 and the value of the cosmological constant $\Lambda$,
then solution to \eqref{static_system} always exist and is unique (cf. Fig.~\ref{static_sys_fig}).
The well-known Einstein dust solution is:
$$
\rho_E=\frac{\Lambda}{4 \pi G}
$$
with the radius:
\begin{equation}
R_E = \frac{1}{\sqrt{\Lambda}}
\end{equation}

In the case of relativistic EOS $p=\rho/3$ we have:
\begin{equation}
\rho_E = \frac{\Lambda}{8 \pi G}
\end{equation}
\begin{equation}
R_E = \frac{\sqrt{3/2}}{\sqrt{\Lambda}}
\end{equation}

Therefore we can see, that size ($R_E$) and total mass of the Einstein static universe:
\begin{equation}
M_E = \rho_E \;V = 2 \pi^2 {R_E}^3 \rho_E
\end{equation}
vary only slightly. Numerically:
\begin{equation}
\label{radius_num}
1/\sqrt{\Lambda}<R_E<1.22 / \sqrt{\Lambda}
\end{equation}
and:
\begin{equation}
\frac{1.57}{G\, \sqrt{\Lambda}}>M_E>\frac{1.44}{G \,\sqrt{\Lambda}}
\end{equation}

If one allows more ''exotic'' matter (e.g. with negative
pressure or density) it is possible to produce solutions
spanning much more wide parameter range, see \cite{1967MNRAS.137...69H}.

It is illustrative to compute $R_E$ and $\rho_E$ using the value of $\Lambda$ recently determined from WMAP and SDSS observations.
 According to \cite{PhysRevD.69.103501} cosmological
constant is:
$$
\frac{\Lambda c^2}{G} = 6.95 \times 10^{-27} kg/m^3
$$
and density of the baryonic and dark matter is 
$\Omega_m+\Omega_b\simeq0.41\, \Omega_{\Lambda}$.
From \eqref{static_system} we get:
$$
\frac{1}{2}\, \Omega_{\Lambda} < \Omega^{\mathrm{s.E.u.}}_{m+b} < \Omega_{\Lambda}.
$$

Radius of the static model \eqref{radius_num} is proportional to:
$$
\frac{1}{\sqrt{\Lambda}} \simeq \text{14.3 Gpc.}
$$

S.E.u. however, once considered very seriously 
\citep{1930MNRAS..90..668E}, is no longer considered a physical reality.

In the next section we would like to present alternative view of the static Einstein universe:
 one of the static spherically symmetric solutions
of the TOV-$\Lambda$ equation with barotropic fluid.

Our goal is to show the existence and explore properties
of the wider than s.E.u. class of  static solutions, namely spherically
symmetric, possibly closed, universes with EOS \eqref{polytropic_fluid}. 
As we will show in the next sections, 3-sphere homogeneous solution is not
an exclusive solution to the given problem. In particular, we discuss 
Tolman-Oppenheimer-Volkov equation with cosmological constant (next section),
generalized Tooper equation with non-zero $\Lambda$ (Sect. \ref{sect:Tooper-lambda})
and additional ''void equation'' \eqref{void_equation} in subsection \ref{subsect:voidEQ}.
Set of both numerical and analytical solutions 
to these equations is then used to fit observed properties of the cosmological voids.

\section{TOV-$\Lambda$ equation }  \label{sect:TOVlambda}

Using standard static spherically symmetric spacetime metric:
\begin{equation}
\label{schwarzchild}
ds^2 = -e^{2 \nu(r) } \; dt^2 + e^{2 \lambda(r) } d r^2 + r^2 \, d \Omega^2
\end{equation}
we can easily derive Einstein field equations \citep{Stuchlik}
for static ideal fluid with barotropic EOS \eqref{barotropic_fluid}.
As these calculations are extensively discussed in literature and textbooks
we jump immediately to final form of the equations:
\begin{subequations}
\label{TOV-lambda-sys}
\begin{equation}
\label{TOV-lambda-eq}
\frac{dp}{dr} = 
- G\;
\frac{ \left( \rho + \frac{p}{c^2} \right) \left( m+ 4 \pi r^3 \, \frac{p}{c^2} - \frac{1}{3}\; \frac{r^3 \Lambda c^2}{G} \right)}
{r^2\; (1- \frac{2\;G m}{c^2 r} - \frac{1}{3}\; r^2 \Lambda)}
\end{equation}
\begin{equation}
\label{TOV-lambda-eq-mass}
\frac{dm}{dr} = 4 \pi r^2 \rho
\end{equation}
\end{subequations}

Eq.~\eqref{TOV-lambda-eq} is referred to as TOV-$\Lambda$
equation. If we put $\Lambda=0$ in \eqref{TOV-lambda-eq}
we get standard TOV equation; if additionally we neglect
pressure as a source of gravitational attraction 
and drop term $1-2m/r$ in  the denominator, responsible 
for spacetime deformation in the presence of strong gravitational
field, we get standard condition of hydrostatic equilibrium
in Newtonian theory. 

If we solve system \eqref{TOV-lambda-sys} then metric \eqref{schwarzchild}
components are easily obtained from formulae:
\begin{equation}
e^{2 \lambda} = \frac{1}{1-2\;m/r-1/3 \; r^2 \Lambda}
\end{equation}

\begin{equation}
\label{nur}
e^{2 \nu} = -\int \frac{dp}{\rho+p} \equiv h(r) 
\end{equation}
where $h$ is the enthalpy.

As an exercise, we would like to look for constant solutions
of the TOV-$\Lambda$ equation. Let us substitute values $p(r)=p_E$ and $\rho(r)=\rho_E$
into \eqref{TOV-lambda-sys}. We get immediately condition
\eqref{static_source}. The $g_{rr}$ component
of the metric can be put in the form:
$$
e^{2 \lambda} = \frac{1}{1-r^2/{R_E}^2}
$$
with:
$$
\frac{1}{{R_E}^2} = \Lambda-8 \pi G p_E.
$$
We have got 3-sphere with radius $R_E$ equal to value determined for 
static Einstein universe \eqref{static_system_2}, as expected.

Very interesting question arise: what happen if we take an initial
value\footnote{Initial value for second equation
\eqref{TOV-lambda-eq-mass}  is $m(r\!=\!0)=0$.
}
for eq.~\ref{TOV-lambda-eq} :
$$
\rho(r=0)\!\equiv\!\rho_0 \neq \rho_E
$$.

\begin{figure*}
\begin{tabular}{cc}
\includegraphics[width=0.5\textwidth]{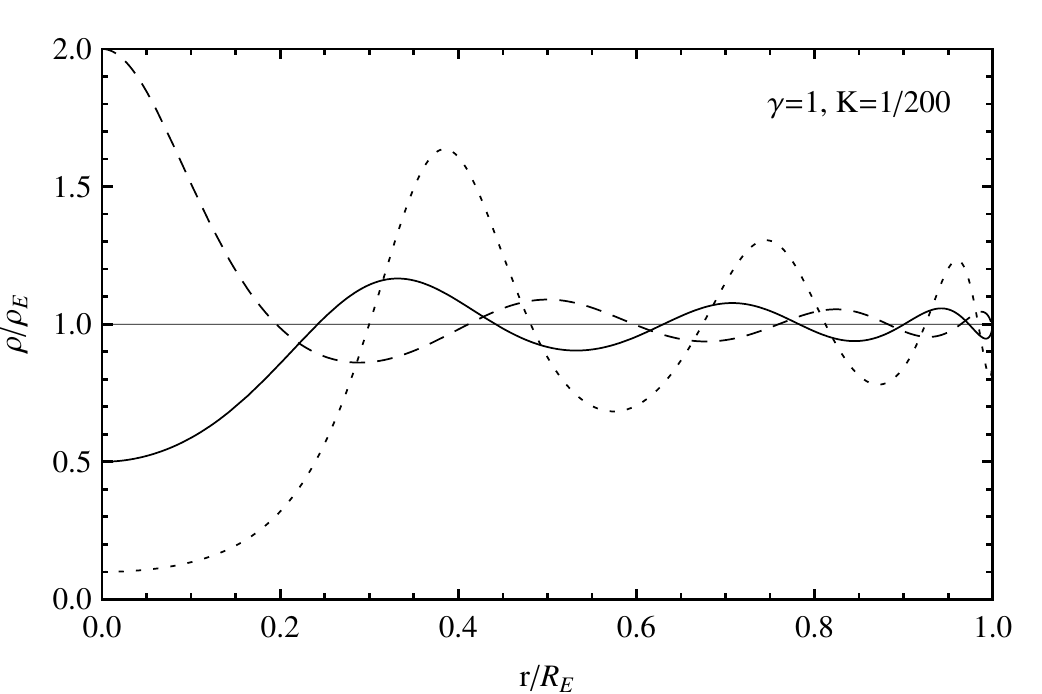}
&
\includegraphics[width=0.5\textwidth]{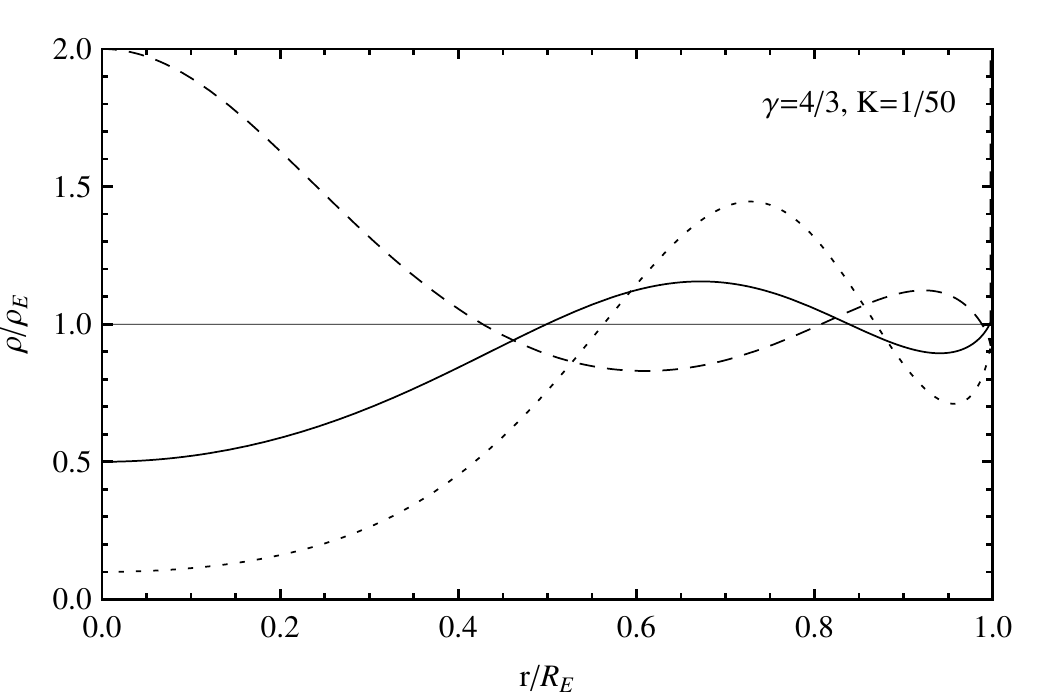}
\\
\includegraphics[width=0.5\textwidth]{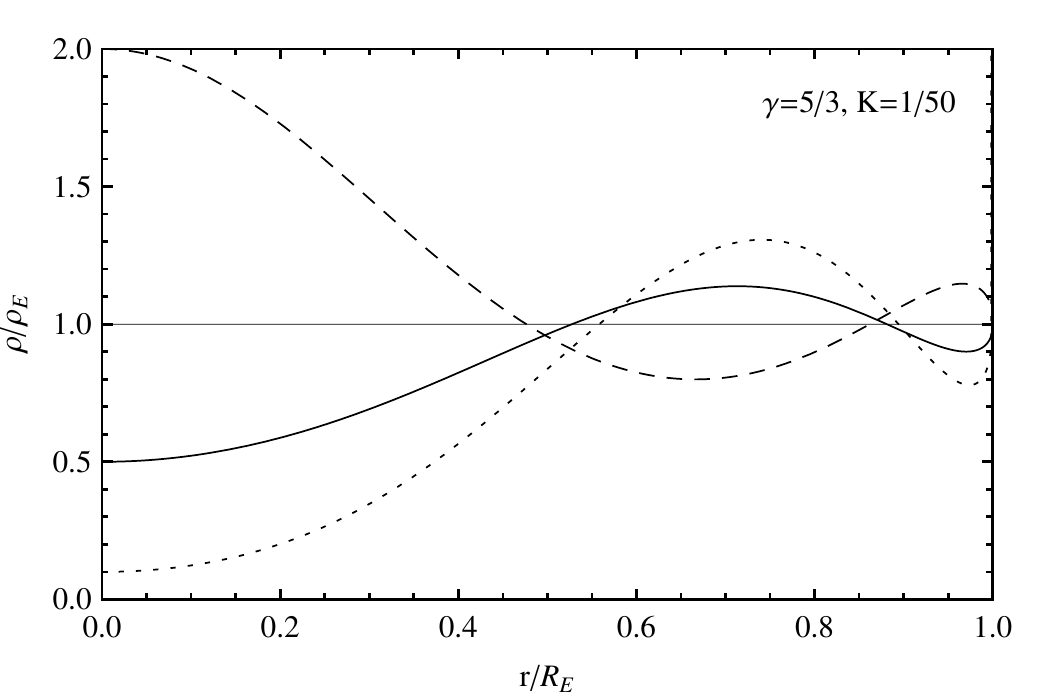}
&
\includegraphics[width=0.5\textwidth]{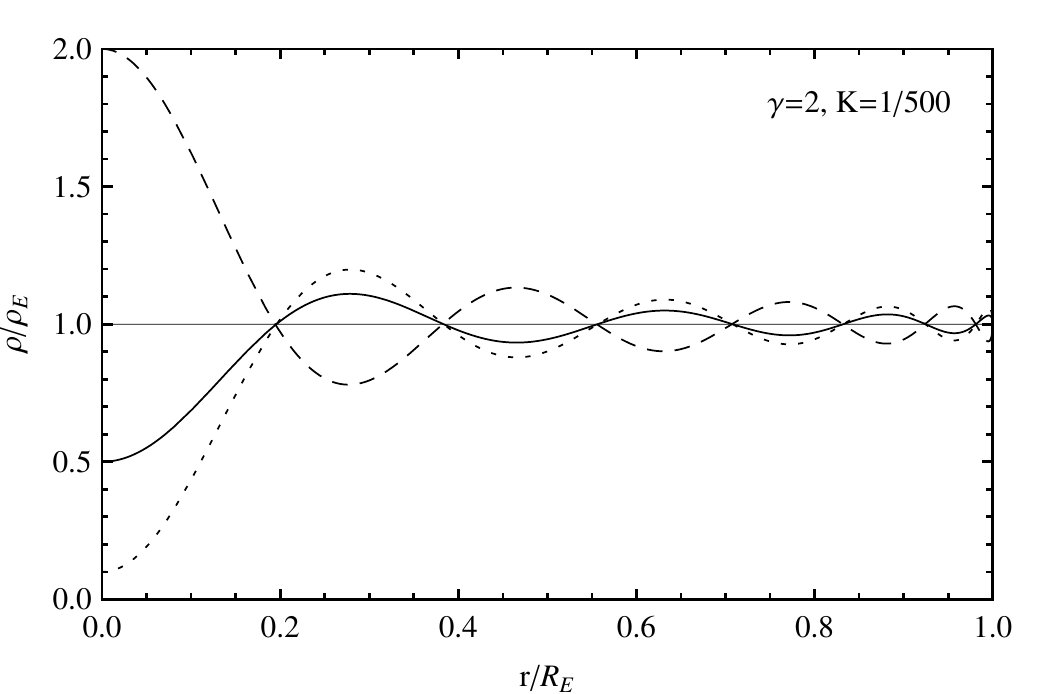}
\end{tabular}
\caption{\label{TOV-lambda-examples} Some selected numerical solutions
to the \eqref{TOV-lambda-sys} with barotropic EOS \eqref{polytropic_fluid}.
Values used to obtain figures are: $\Lambda=4 \pi$, $G=1$; respective
 density and radius of the dust s.E.u are $R_E= 1 / \sqrt{4 \pi} $ and $\rho_E=1$.
Initial values are: $\rho_0=\rho_E/10$ (dotted) and $\rho_E/2$ (solid)
for a ''hole'' and $2 \rho_E$ for ''overdensity region'' (dashed).}
\end{figure*}

The problem is solvable numerically.
However, removable singularity
is present at $r=0$. To start numerical integration of \eqref{TOV-lambda-sys}
we follow standard procedure, replacing unknown functions with a series
expansion:
\begin{subequations}
\begin{equation}
\rho(r) \simeq \rho_0 + \alpha\, r + \frac{1}{2} \beta\, r^2 + \ldots
\end{equation}
\begin{equation}
p(r) \simeq p_0 + a\, r + \frac{1}{2} b\, r^2 + \ldots
\end{equation}
\begin{equation}
m(r) \simeq m_0 + A\, r + \frac{1}{2} B\, r^2 + \frac{1}{6}  C\, r^3 + \ldots.
\end{equation}
\end{subequations}
Substituting series expansion into \eqref{TOV-lambda-sys} immediately
gives:
\begin{subequations}
\begin{equation}
\rho(r) \simeq \rho_0  + \frac{1}{2} \beta r^2 + \ldots
\end{equation}
\begin{equation}
p(r) \simeq p_0 +  \frac{1}{2} b r^2 + \ldots
\end{equation}
\begin{equation}
m(r) \simeq \frac{4}{3} \pi r^3 \rho_0 + \ldots.
\end{equation}
\end{subequations}
where:
\begin{subequations}
\label{TOV-lambda-series}
\begin{equation}
b=\frac{4}{3} \pi G (p_0 + \rho_0) (\rho_E+3 p_E - \rho_0 - 3 p_0 )
\end{equation}
and:
\begin{equation}
\beta = b / \left ( \frac{\partial p}{\partial \rho} \right )_{\rho=\rho_0}.
\end{equation}
\end{subequations}

From \eqref{TOV-lambda-series} we can distinguish three classes of the solutions:
\begin{enumerate}
\item[(A)]{If $\rho_E+3 p_E < \rho_0+3 p_0$ then $b<0$ and we get solution
similar to normal ''star'' (cluster, etc.) type solutions where density and pressure 
decreases outwards.}
\item[(B)]{$\rho_E+3 p_E > \rho_0+3 p_0$ implies $b>0$ i.e. both 
pressure and density 
grow with radius. This solutions describes ''hole'' in the background matter.
}
\end{enumerate}

Finally, if $\rho_E+3 p_E = \rho_0 + 3 p_0$ we get static Einstein universe,
boundary solution between ''clusters'' and ''voids''.

Because class (B) of the solutions remained unnoticed\footnote{ 
In dimensionless form, eq.~\eqref{TOV-lambda-dimensionless},
it is much easier to see the existence of the two solution classes, cf. Sect.~\ref{sect:Tooper-lambda}.
} 
(to my knowledge), 
comment is required. The TOV-$\Lambda$ equation  usually is viewed as a perturbation to normal TOV
equation. 
It is applied to the objects like stars. Typical initial value of density at r=0
exceeds that of the s.E.u. by many of orders of magnitude.
Moreover, the integration is 
finished at point where $p \!=\! \rho \!=\! 0$. Therefore, no asymtotic density behavior 
is observed. Additionally, asymptotic behavior depends
on polytropic, like for Lane-Emden equation. For fractional $n$
solution cannot be extendended beyond $\rho<0$. 
If the initial value of source density ($\rho+3\, p$) is however comparable to
$\rho_E+3 p_E$ (but still in the regime (A) ), then the new behavior is observed. Instead of going to $\rho=0$, density
approaches constant value, exhibiting oscillatory 
behavior, cf. Fig.~\ref{TOV-lambda-examples}. It happens for relatively 
small subset of all possible initial conditions.

Surprisingly, starting the integration of \eqref{TOV-lambda-sys} with 
central density in the regime (B), i.e. less than $\rho_E+3\, p_E$, 
we get qualitatively different
solution. We get family of static, underdensity objects, again approaching constant
density (Fig.~\ref{TOV-lambda-examples}). Analysis of the original TOV-$\Lambda$
with general EOS \eqref{barotropic_fluid} is very difficult task. Therefore,
we have concentrated on its dimensionless \eqref{TOV-lambda-dimensionless}
and simplified \eqref{void_equation} forms. They are valid for polytropic EOS \eqref{polytropic_fluid}.

We will show in the Section \ref{sect:HoleProperties} that the existence of these objects is a manifestation
of the repulsive gravity related to the non-zero value of the cosmological constant.

\section{Properties of the void model in the form of the ''hole'' }\label{sect:HoleProperties}

''Holes'' (Fig.~\ref{TOV-lambda-examples}) are very exotic objects. Density
{\em increases} outwards and pressure gradient is {\em positive}. Newtonian
intuition tells us, that matter inside given radius should
attract particle outside. It is however simple to show, that test
particle will be repelled from the 
hole\footnote{Or attracted by the universe outside. Unfortunately, 
for the closed universe there is no clear method to define
which part  of the 3-sphere split by 2-sphere into two regions,  is ''outside'' or ''inside'' 2-sphere. 
We remind, that this solution is not asymptotically flat, and describe spatially finite space
}
using relativistic approach.

\begin{figure*}
 \begin{tabular}{cc}
\includegraphics[width=0.5\textwidth]{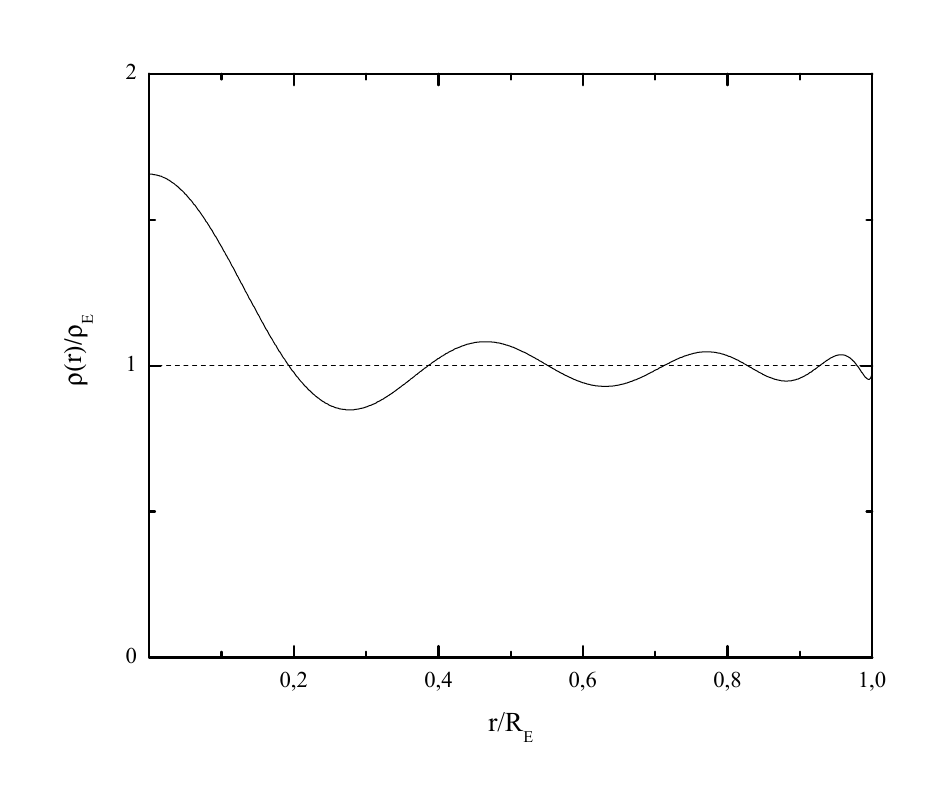}
&
\includegraphics[width=0.5\textwidth]{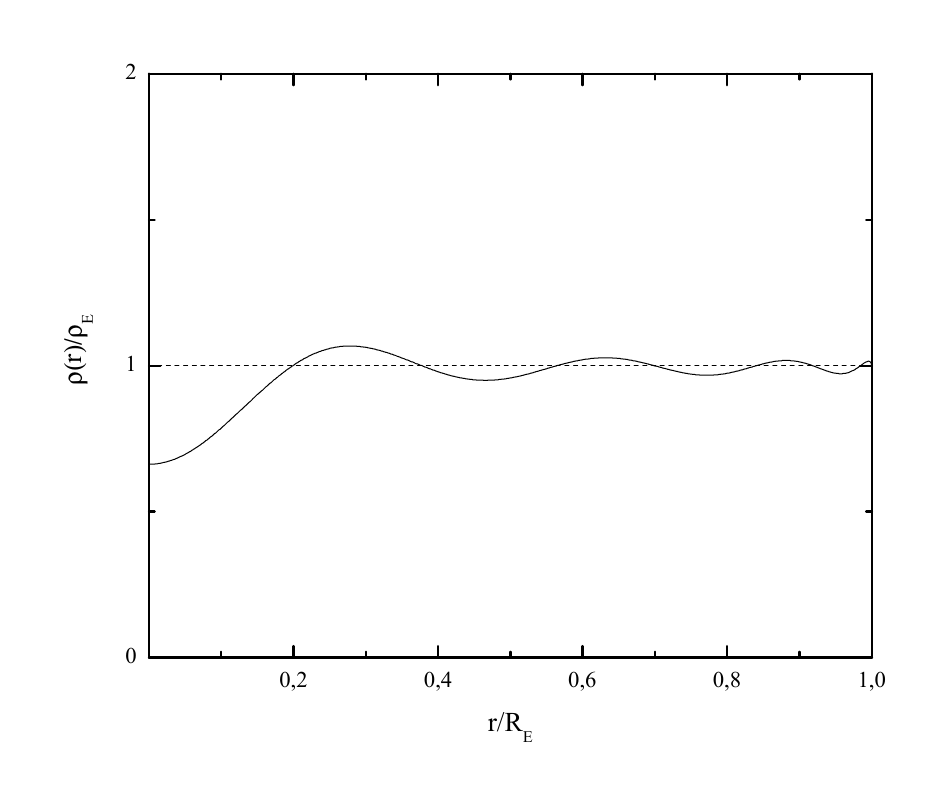}\\

\includegraphics[width=0.5\textwidth]{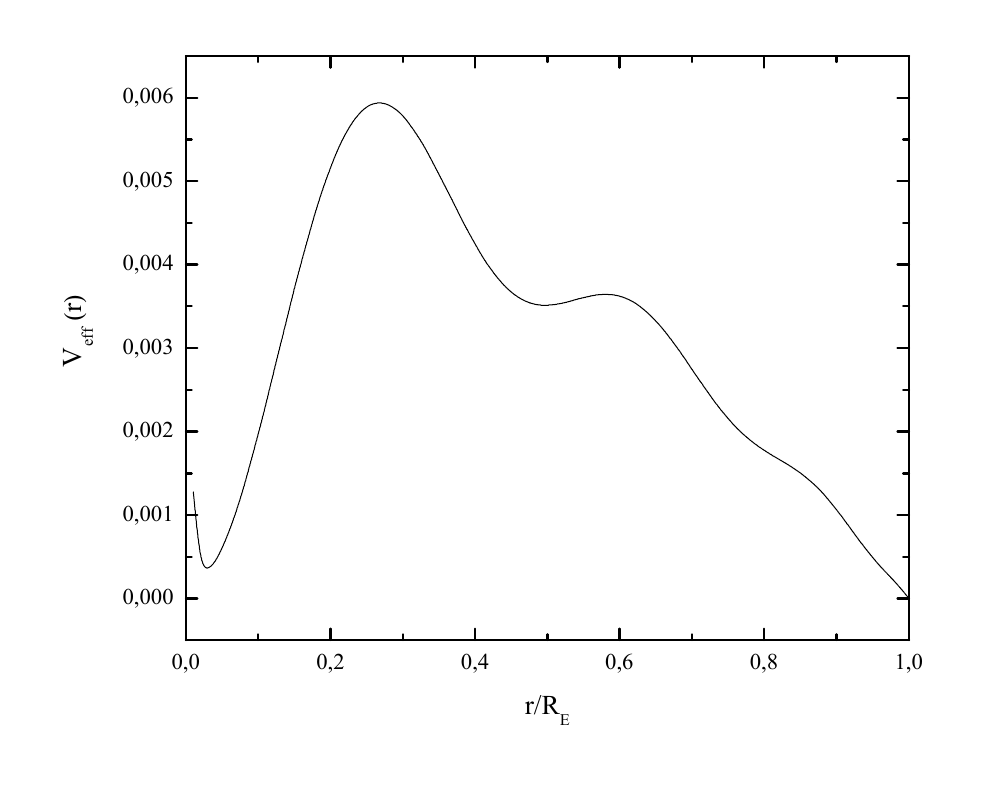}
&
\includegraphics[width=0.5\textwidth]{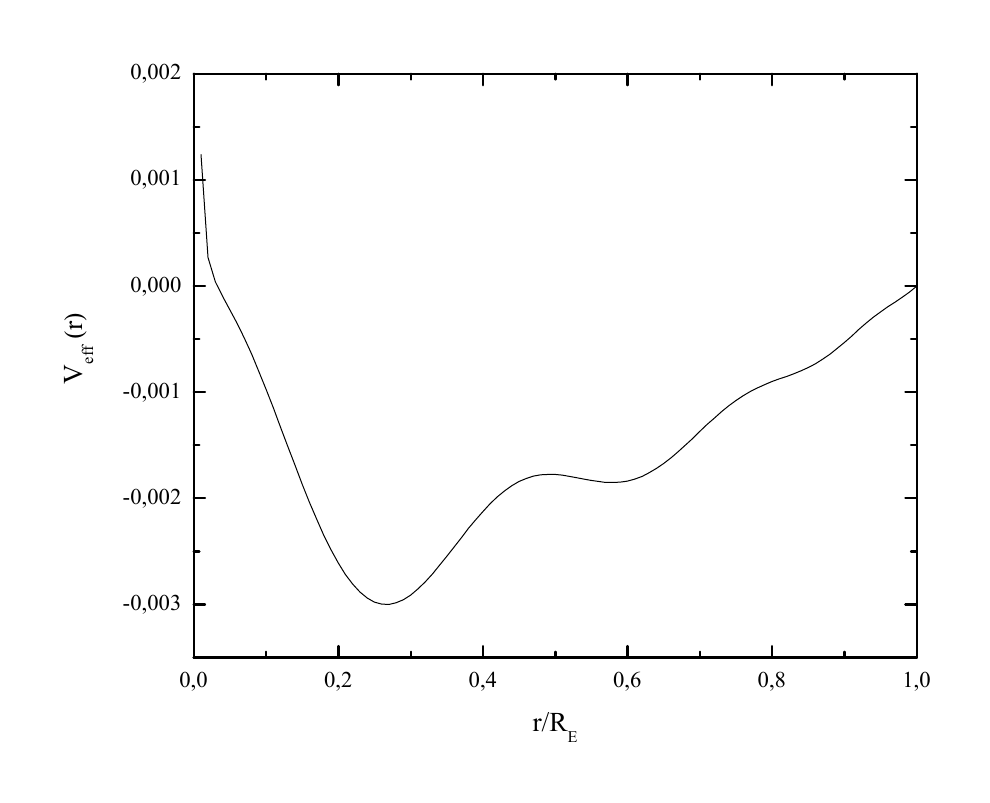}
\end{tabular}
\caption{\label{Veff} Density profile for overdensity object (upper left)
and hole (upper right) with effective potential \eqref{eff_pot} with 
very small $L$ below. Stable circular orbit exist very close
to the center of overdensity region (lower left), while extremum do not exist
inside a hole for identical $L$ and $E$. Stable orbit, however, exist
 very close to the ''surface'' of the hole. This behavior is easy to
 understand if we note, that source of gravitational attraction/repulsion
 is over/under--density relative to values for s.E.u., not entire matter
inside a given radius.    
}
\end{figure*}

Because our solution is spherically symmetric we may use the effective potential method:
\begin{equation}
\label{eff_pot}
V_{eff} = \Bigg ( 1- \frac{2 m}{r} - \frac{1}{3} \Lambda r^2
\Bigg ) \Bigg (1+\frac{L^2}{r^2} - \frac{E^2}{e^{2 \nu}} \Bigg)
\end{equation}
where $m(r)$ and $\nu(r)$ are computed from \eqref{TOV-lambda-eq-mass} 
and \eqref{nur} using numerical solution (Fig.~\ref{Veff}). If no local extremum exist,
then circular orbits (both stable and unstable) 
are impossible and ''force'' is repulsive. This is the case
of the hole (cf. Fig.~\ref{Veff}, right panel). Therefore, 
we have shown, that the hole
repels test particles\footnote{Thus, intriguing question arise
about ''force'' between two holes.}
from inside. Without positive pressure gradient
it cannot be static. This also explains why non-trivial 
EOS, e.g. \eqref{polytropic_fluid}, is required for this type of the solution.  
Dust particles will be repelled from the hole. Static Einstein universe
is an exception, because of its high symmetry.

\section{Generalized Tooper fluid spheres and dimensionless 
form of the TOV-$\Lambda$ equation} \label{sect:Tooper-lambda}

By generalization of the  \citet{1965ApJ...142.1541T} result, 
we can put TOV-$\Lambda$ equations \eqref{TOV-lambda-sys}
in dimensionless form. This is possible because power-law equation of state \eqref{polytropic_fluid} 
has been used. For general barotropic EOS, eq.~\eqref{TOV-lambda-sys} must
 be used instead.

Let us introduce new dimensionless radial variable:
$$
x = A \, r,
$$
and new functions:
\begin{subequations}
\begin{equation}
\label{densiy_profile}
\rho(r) = \rho_c\, \theta(x)^n,
\end{equation}
\begin{equation}
p(r) = p_c\, \theta(x)^{n+1}, 
\end{equation}
\begin{equation}
m(r) = \frac{4 \pi \rho_c}{A^3} v(x).
\end{equation}
\end{subequations}
Here, $\rho_c$ is the density at the center ($r=0$) and $p_c$ is the central pressure.

Now, Eqns.~\eqref{TOV-lambda-sys} take the form:
\begin{subequations}
\label{TOV-lambda-dimensionless}
\begin{equation}
\label{TOV-lambda-dimensionless1}
\theta '+
\frac{ (1 + \alpha \theta) (v+\alpha x^3 \, \theta^{n+1} - \lambda\, x^3/\alpha) }
{x^2 - 2 \alpha\, x\, v- \lambda\, x^4}
=0
\end{equation}
\begin{equation}
v '= x^2\, \theta^n
\end{equation}
\end{subequations}
where:
$$
A^2=\frac{4 \pi G \rho_c}{\alpha c^2}, 
\quad 
\lambda = \frac{\Lambda}{3 \, A^2}, 
\quad
\alpha=\frac{p_c}{\rho_c c^2}, \quad
\gamma = 1 + \frac{1}{n}.
$$
Initial conditions for system \eqref{TOV-lambda-dimensionless} are: 
$\theta(0)=1$ and $v(0)=0$. If cosmological constant (thus parameter $\lambda$)
is equal to zero, and we neglect relativity parameter $\alpha$, system \eqref{TOV-lambda-dimensionless}
reduces to the Lane-Emden equation. However, if $\lambda \neq 0$, there is no simple Newtonian
limit. The last term of the numerator in eq.~\eqref{TOV-lambda-dimensionless1} diverge.
We will use this property in subsection \ref{subsect:voidEQ} to build the simple void model.

Behavior of the solution around $x=0$ can be explored using series expansion. Repeating procedure used
previously for \eqref{TOV-lambda-sys}, we get:
\begin{subequations}
\label{Tooper-lambda-series}
\begin{equation}
\label{Tooper-lambda-series1}
\theta(x) \simeq 1 + \frac{1+\alpha}{2} \left( \lambda/\alpha - \frac{1}{3} - \alpha \right)\, x^2 + \ldots
\end{equation}
\begin{equation}
\label{Tooper-lambda-series2}
v(x) = \frac{1}{3}\, x^3 + \ldots
\end{equation}
\end{subequations}

Let us note that constant solution $\theta(x)=1$ (static Einstein universe)
appears only if:
\begin{equation}
\label{lambda_critical}
\lambda = \alpha (\alpha + 1/3) \equiv \lambda_c.  
\end{equation}
Using \eqref{lambda_critical} we rewrite \eqref{Tooper-lambda-series1} in the form:
\begin{equation}
\label{Tooper-lambda-series-alternative}
\theta(x) \simeq 1 + \frac{1+\alpha}{2 \alpha} \left( \lambda - \lambda_c \right)\, x^2 + \ldots.
\end{equation}
Eq.~\ref{Tooper-lambda-series-alternative} shows that $\theta(x)$ decreases only if 
$\lambda < \lambda_c$; $\lambda=0$ in particular.

The transformation from eqns.~\eqref{TOV-lambda-sys} to \eqref{TOV-lambda-dimensionless} 
clarify the number of the independent free parameters in \eqref{TOV-lambda-sys}. 
Density profile
can be obtained from three-parameter special function $\theta_n(x;\alpha, \lambda)$
using \eqref{densiy_profile}.

\section{Fitting ''hole'' model to observed profiles of the cosmological voids }\label{sect:TOV-lambda-fit}

Radial density profile of the voids can be determined from astronomical observations. We have used
data from Fig.~4 of \cite{2004ApJ...607..751H} to compare static model with observations.

\begin{figure*}
\begin{tabular}{cc}
\includegraphics[width=0.5\textwidth]{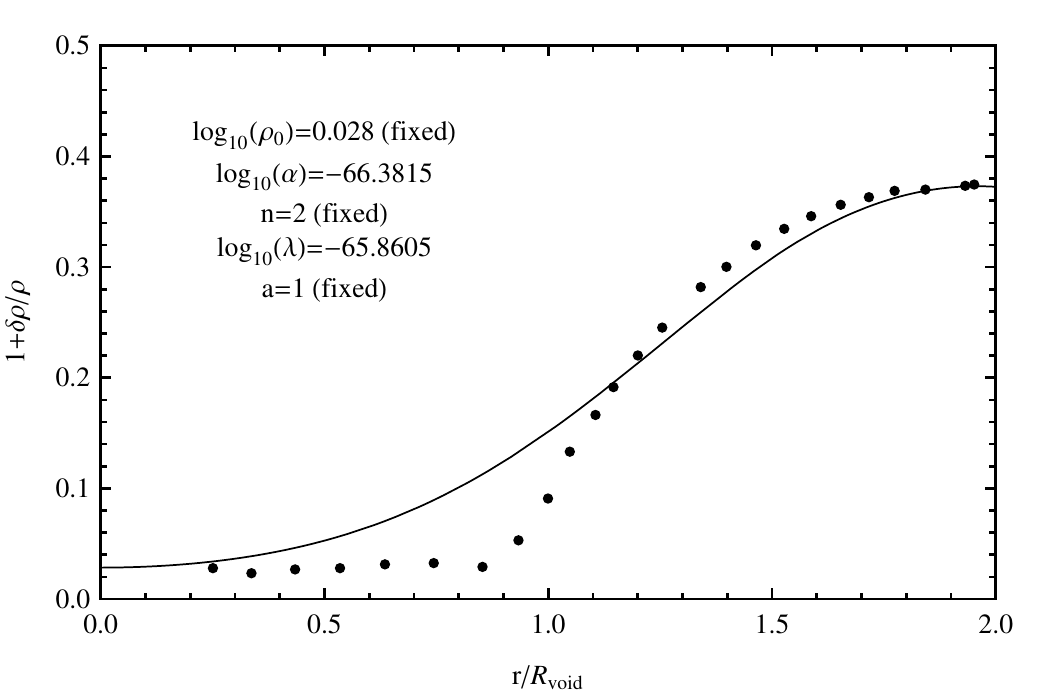}
&
\includegraphics[width=0.5\textwidth]{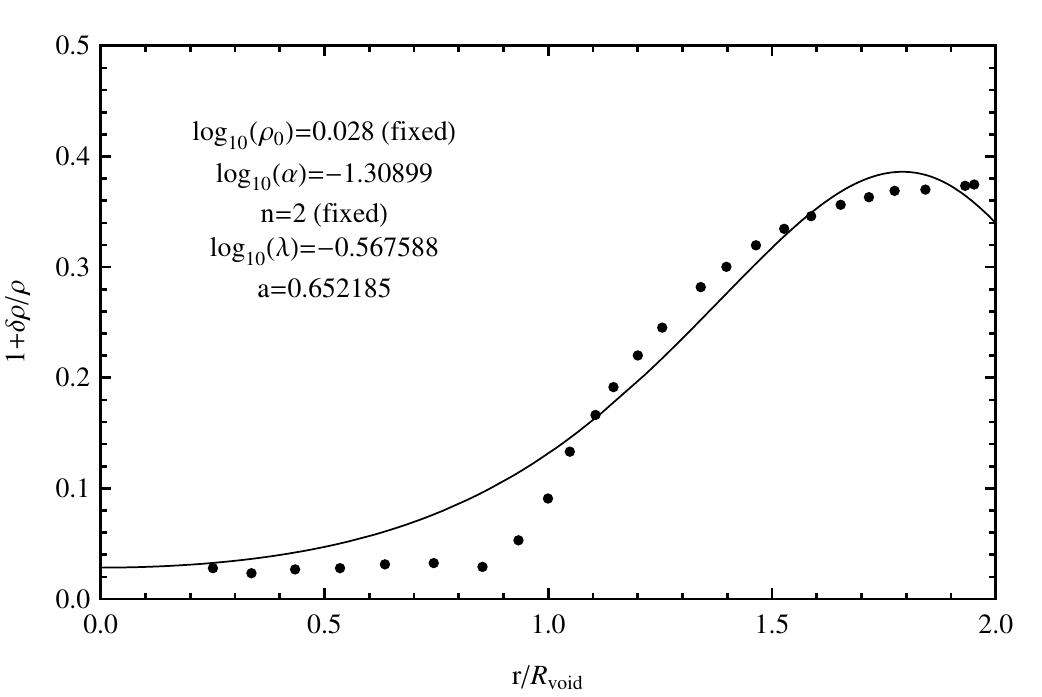}\\
\includegraphics[width=0.5\textwidth]{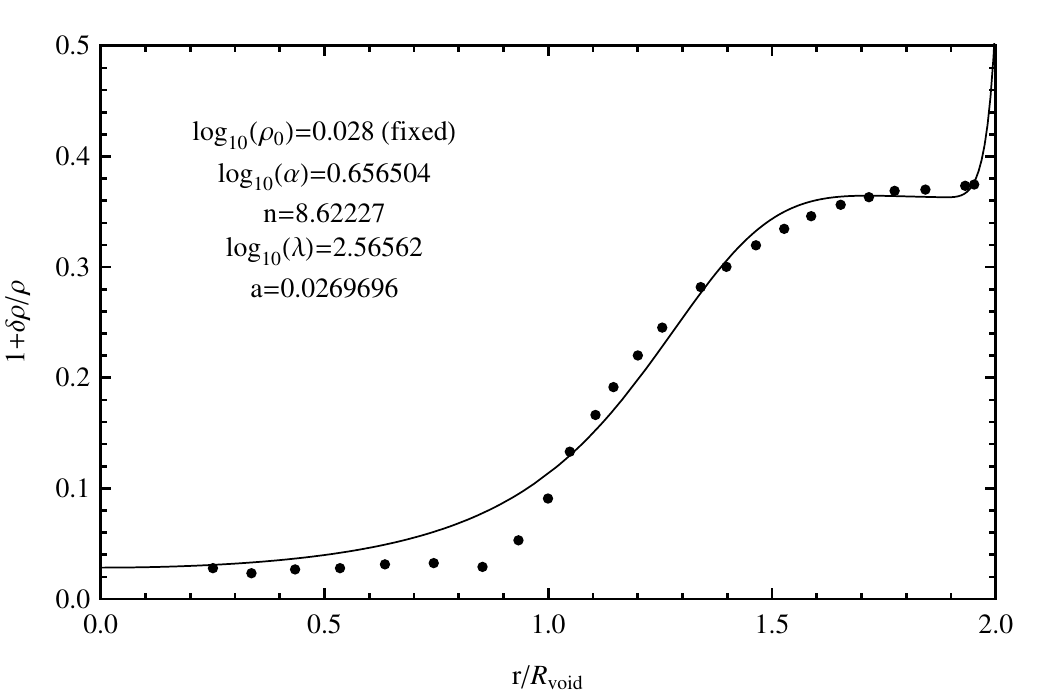}
&
\includegraphics[width=0.5\textwidth]{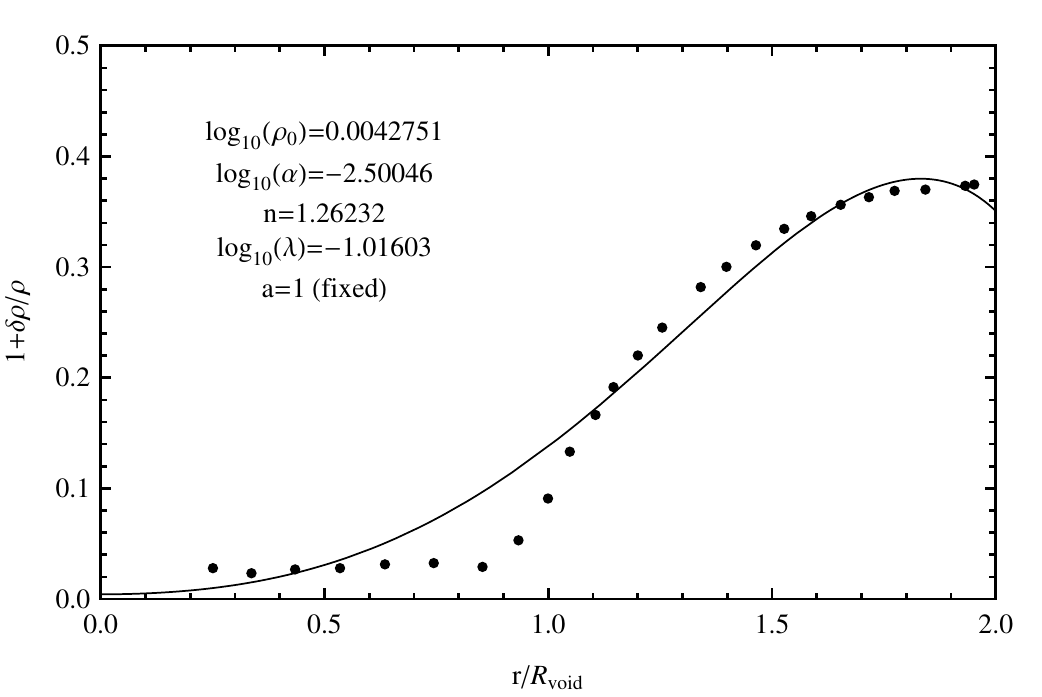}
\end{tabular}
\caption{\label{fits} Best fit solution of the Tooper-$\Lambda$
\eqref{general_fit} and radial profile of void provided by \citep{2004ApJ...607..751H}.
Points show data extracted from ''SGP'' dataset.    
}
\end{figure*}

 In general, static model based on Tooper-$\Lambda$ equation \eqref{TOV-lambda-dimensionless}
 provide 4-parameter family of solutions. If EOS is given
 completely, i.e. without any of the free 
parameters\footnote{For example, if $p=\rho/3$ then $\alpha=1/3$ and $n=\infty$.
}, 
then solution is determined by just two parameters: central density and $\Lambda$.
Solutions are enumerated by the central density $\rho_0$ and value
 of the cosmological constant $\Lambda$. In principle, value of $\Lambda$
 is known from e.g. $\Lambda$-CDM model. However, real voids are 
 not placed in the static Einstein universe, but in $\Lambda$-CDM Universe. 
Some effective $\Lambda_{eff} \neq \Lambda_{CDM}$
 could be used instead. Therefore we consider $\lambda$ in 
\eqref{TOV-lambda-dimensionless} a free parameter. 
Radial profile of the void in \cite{2004ApJ...607..751H} is normalized arbitrarily
 in radius. Therefore additional free parameter $A$ must be used to 
scale solution of the TOV-$\Lambda$
equation.  Polytropic EOS \eqref{polytropic_fluid}
  has been used in fitting with additional parameter, $n=1/(\gamma-1)$.

The most general fit: 
\begin{equation}
\label{general_fit}
\rho(r) = \rho_c \, \theta_n(A r;\, \alpha, \lambda)
\end{equation}
includes five free parameters: central density of the void, $\rho_c$,
constant $n$ coming from polytropic EOS (Eq.~\ref{polytropic_fluid}), the relativity
parameter $\alpha$, the cosmological constant related parameter $\lambda$ and one more parameter, 
$A$, used to scale solution of the TOV-$\Lambda$ equation.

Representative fits are shown in Fig.~\ref{fits}. Overall quality of approximation is good.
Four cases are presented. Because it is unclear why the data points
are missing in Fig.~4 of \cite{2004ApJ...607..751H} for $r<0.2 R_{void}$, we consider two scenarios.
In first, we use fixed central density. The value is determined by extrapolation
of data from $0.25\, R_{void}<r<0.85\, R_{void}$ to $r=0$. Linear fit gives:
$$
\rho(r) = p r +q
$$
where:
$$
p=0.009\pm0.004, \quad q=0.023 \pm 0.003 \quad (1 \sigma).
$$
This is consistent ($p \in (-0.0023, 0.021)$, at 95\% confidence level) with constant density ($p=0$) for $r<0.85 R_{void}$. Assuming
constant density we get:
$$
\rho_c = 0.028 \pm 0.001 \quad (1 \sigma),
$$
and this value has been used in fits.

Four typical cases are presented in Fig.~\ref{fits}. Upper-right panel
of Fig.~\ref{fits} will be examined in greater detail in subsection~\ref{subsect:voidEQ}.
All cases are summarized in Table~\ref{fitTBL}. Except for lower-left figure in Fig.~\ref{fits},
$\alpha \ll 1$ and $\lambda \ll 1$. Little can be told on the polytropic index $n$.
For all four cases $1<n<9$. The best fit has been obtained with fixed central density (lower-left panel
in Fig.~\ref{fits} ). Value of $\lambda$ is very large in this case, and polytropic index $n>8$.
This suggest that voids are $\Lambda$-dominated objects filled with ideal
gas with a lot ($2 n>16$) of the degrees of freedom. This, unfortunately,  
contradicts results of subsection~\ref{subsect:Rrho}. However, used density profile
is the averaged one. Profiles for individual voids would work better,
but there is no method to measure them. Too many degrees of freedom
in model \eqref{general_fit} combined with limited amount of data allow us
to provide hints rather than constraints on EOS. 

\begin{table}
\caption{\label{fitTBL} Parameters used in Fig.~\ref{fits}.}
\begin{tabular}{cccccc}
$\rho_c$ & n & $\log_{10} \alpha$ &$\log_{10} \lambda$ & A \\
\hline
\hline
0.028 (fixed) & 2 (fixed) & -66.38 & -65.86 & 1 (fixed) \\
0.028 (fixed) & 2 (fixed) & -1.3 & -0.57 & 0.65 \\
0.028 (fixed)  &  8.6 & 0.66 & 2.6 & 0.027\\
0.004 $\pm$ 0.015 & 1.26 $\pm$ 0.29 & -2.5 $\pm$ 1.5 & -1.0 $\pm$ 0.2 & 1 (fixed)  
\end{tabular}
\end{table}

\subsection{The ''void equation''} \label{subsect:voidEQ}

\begin{figure}
\includegraphics[width=0.5\textwidth]{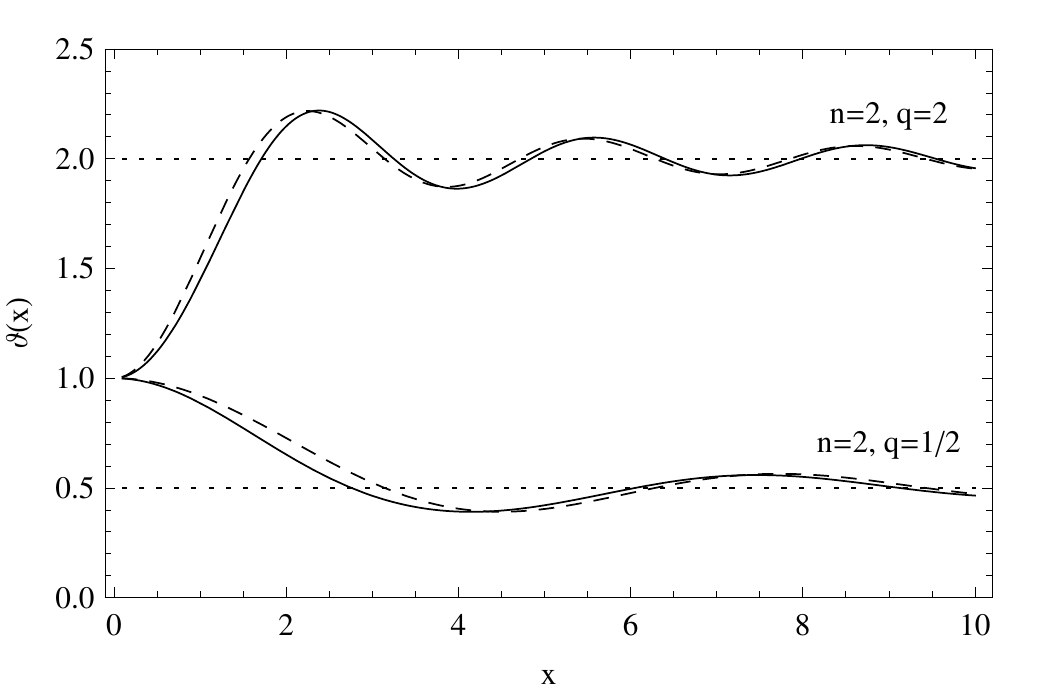}
\caption{\label{approx_fig} Example solutions to the ''void equation'' \eqref{void_equation} (solid line)
and approximate analytical formula \eqref{void_eq_approx} (dashed line).
Dotted horizontal lines show asymptotic value $\lim_{x \to \infty} \vartheta_n(x)=q$.
}
\end{figure}

The case with fixed
polytropic index, $\gamma=2$ and fixed central density
(Fig.~\ref{fits}, upper-left) is particularly intriguing. 
Solution is not scaled in radius, i.e $A=1$. We have just two free parameters:
$\alpha$ and $\lambda$. Best fit parameters (upper-left panel in Fig.~\ref{fits})
are very small, and are poorly determined. Actually, from least-square
fits we get random pairs of extremely small values. However, ratio $\lambda/\alpha$
is almost constant. In this situation we can simplify eqns.~\eqref{TOV-lambda-dimensionless}
a lot. Assuming that $\alpha \to 0$ and $\lambda \to 0$ and $\lambda/\alpha \to const$ we get:
\begin{subequations}
\begin{equation}
\theta'+\frac{v- p x^3}{x^2}=0
\end{equation}
\begin{equation}
v'=x^2 \theta^n
\end{equation}
\end{subequations}
where we denoted $p=\lambda/\alpha$.
For $n=1$ (initial conditions the same like for \eqref{TOV-lambda-dimensionless}) 
we get analytical solution for the density profile of the void:
$$
\rho(r) = \rho_c \; \frac{\sin{x}-3 p \sin{x} + 3 p x}{x}, \quad x=A r.
$$
For $n \neq 1$ we still have to use numerics (see however 
eq.~\ref{void_eq_approx} and Fig.~\ref{approx_fig}),
but the basic equation is much simpler (compared to \eqref{TOV-lambda-dimensionless}):
\begin{equation}
\label{void_equation}
\vartheta'' + \frac{2}{x} \vartheta' +  \vartheta^n = 3 p \equiv q^n, \quad \vartheta(0)=1, \vartheta'(0)=0.
\end{equation}

For $q=0$ eq.~\eqref{void_equation} reduces to the Lane-Emden equation.
For $q \neq 0$ behavior of the solutions\footnote{
Scaling symmetry of the Lane-Emden equation 
$\zeta^{s} \vartheta(x/\zeta)$, where $s=2/(1-n)$
is no longer valid if $q \neq 0$ in \eqref{void_equation}.
} 
for both $x \to 0$ and $x \to \infty$ changes dramatically. Here we restrict to
positive solutions only: $\vartheta>0$, $q>0$ and $n>0$.

For $x \ll 1$ we have:
$$
\vartheta(x) \simeq 1 + \frac{q^n-1}{6} \, x^2 + \ldots,
$$
and depending on sign of $q^n-1$ we have constant, increasing or decreasing solution.
For $x \to \infty$ we have obtained asymptotic expression:
\begin{equation}
\label{void_eq_asymptotic}
\vartheta(x) \sim q - \frac{a \; \sin{ \omega x }}{x},
\end{equation}
where:
\begin{equation}
\label{omega}
\omega^2 = n \, q^{n-1}.
\end{equation}
The constant $a$ can be determined from numerical solution. Expression
\eqref{void_eq_asymptotic} provides surprisingly good (see Fig.~\ref{approx_fig}) overall fit to the
solution in the entire interval $[0,\infty]$. Value of the constant $a$
in \eqref{void_eq_asymptotic} can be roughly determined from
initial condition $\vartheta(0)=1$. This gives $a=(q-1)/\omega$:
\begin{equation}
\label{void_eq_approx}
\vartheta(x) \simeq q - (q-1) \; \frac{  \sin{ \omega x }}{\omega x}.
\end{equation}
Expression \eqref{void_eq_approx} can be used instead of numerical
solution for \eqref{void_equation} if large accuracy is not required.
Maximum relative error is never larger than 15\% for $t>0$, and is 
below 7\% for $1<n<5$ and $0<q<2$.

The most important
in our application (a void model) is the leading constant in \eqref{void_eq_asymptotic}.
Knowledge of the behavior at infinity allow us 
(in contrast to eqns.~(\ref{TOV-lambda-sys}, \ref{TOV-lambda-dimensionless}) )
to introduce the {\em density contrast}:
\begin{equation}
\frac{\delta \rho}{\rho} \equiv \frac{\rho - \rho_{\infty}}{\rho_{\infty}  }.
\end{equation}
Using:
$$
\rho(r) = \rho_c \, \vartheta( A x)^n, \quad \rho_{\infty} = \rho_c \, q^n
$$
we have:
\begin{equation}
\label{densitycontrast_fit}
1 + \frac{\delta \rho}{\rho} = \left ( \frac{\vartheta( A x)}{q} \right)^n
\end{equation}
where $\vartheta$ is solution to the eq.~\eqref{void_equation} or
its approximation \eqref{void_eq_approx}.

We have obtained void model much simpler and convenient
compared to original function \eqref{general_fit}
which must be evaluated numerically. Using \eqref{densitycontrast_fit}
and \eqref{void_eq_approx} we can fit density contrast using three parameters.
Two constants: $A$ and $\omega$ always appear as a product $A \omega$, and
should be treated like a single parameter.

\begin{figure}
\includegraphics[width=0.5\textwidth]{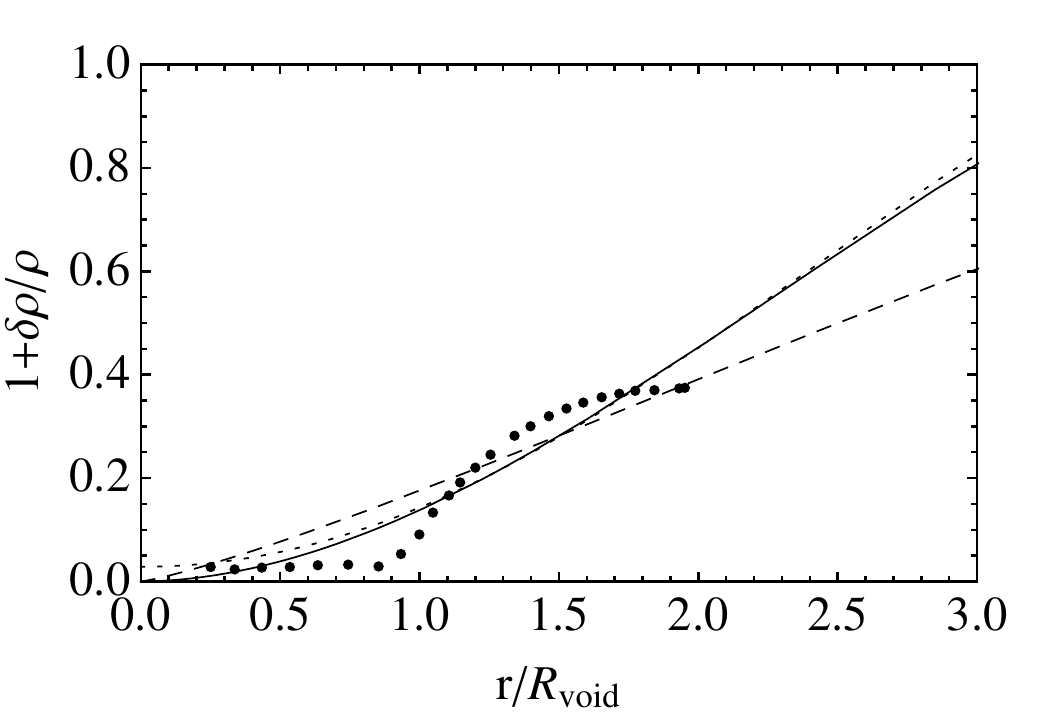}
\caption{ \label{fit_anal} Fitting results obtained with aid of the analytical formula 
(\ref{densitycontrast_fit}, \ref{void_eq_approx}) for the density contrast in the void.
See the text for the details.
}
\end{figure}

Typical fitting results are shown in Fig.~\ref{fit_anal}. Quality
is apparently lower compared to the Tooper-$\Lambda$ fits (Fig.~\ref{fits}).
But curves in Fig.~\ref{fit_anal} represent functions with
well-defined and correct asymptotic behavior. Probably, the polytropic
EOS \eqref{polytropic_fluid} is too simple to reproduce fine structure
of the density profile. Obtained values of the polytropic index $n$
also represent some average value.

Three curves in Fig.~\ref{fit_anal} are presented. Solid line shows unconstrained fit.
Obtained values are: $n=0.9 \pm 0.2$, $A \omega = 0.86 \pm 0.09$ and $q \to \infty$.
Because $\rho_c \propto 1/q^n$ we have obtained void model that is empty
at the very center. Similar result ($n=3/5$, $A \omega = 0.58 \pm 0.02$ and $q \to \infty$)
has been obtained (Fig.~\ref{fit_anal}, dashed line) with fixed polytropic index of $n=3/5$. 
If we fix the central
density:
$$
1/q^n = \rho_c/\rho_\infty
$$
then $n=1.05 \pm 0.2$ and $A \omega = 0.89 \pm 0.09$, see Fig.~\ref{fit_anal}, dotted line.

This simplest model do not produce impressive results,
and little information on the matter filling the void can be extracted. Unfortunately,
data is missing in crucial region $r<0.2 R_{void}$, as well as for $r>2 R_{void}$.
Extrapolating data we expect $\rho_0 \simeq 0.028$ and $\rho_0 \to 1$ for $r \gg R_{void}$. 
Sudden change of derivative at $r \simeq R_{void}$ strongly suggest
different EOS, more complicated than \eqref{polytropic_fluid}.

\subsection{Radius-central density relation for the voids} \label{subsect:Rrho}

In addition to the void density contrast profile, $D_{void} - \rho_c$ relation has been published in
\cite{1995A&A...301..329L}. We can use \eqref{void_eq_approx} to derive this relation.
We define void radius as a point where density contrast becomes equal to 1 for the first time.
Therefore, problem reduces simply to the first zero of the $\sin{\omega x}/(\omega x)$, i.e:
$$
A\omega\, R_{void} = \pi.
$$
Using $\omega$ from \eqref{omega} and $1/q^n = \delta \rho_c$, we get:
\begin{equation}
\label{Rvoid_vs_rhoc}
R_{void} = \frac{\pi}{\sqrt{n} A} \; \delta \rho_c^{1-\gamma/2} = \frac{\pi}{\sqrt{n} A} \; \delta\rho_c^{\frac{n-1}{2 n}}.
\end{equation}
Here, $\delta \rho_c $ is the central density contrast plus 1. Using $A$ and $\gamma$(=$1+1/n$)
as a free fit parameters, we get:
$$
A = 0.23 \pm 0.01, \quad \gamma=2.66 \pm 0.07 \quad \text{i.e.} \quad  n= 0.602 \pm 0.025
$$ 

\begin{figure}
\includegraphics[width=0.5\textwidth]{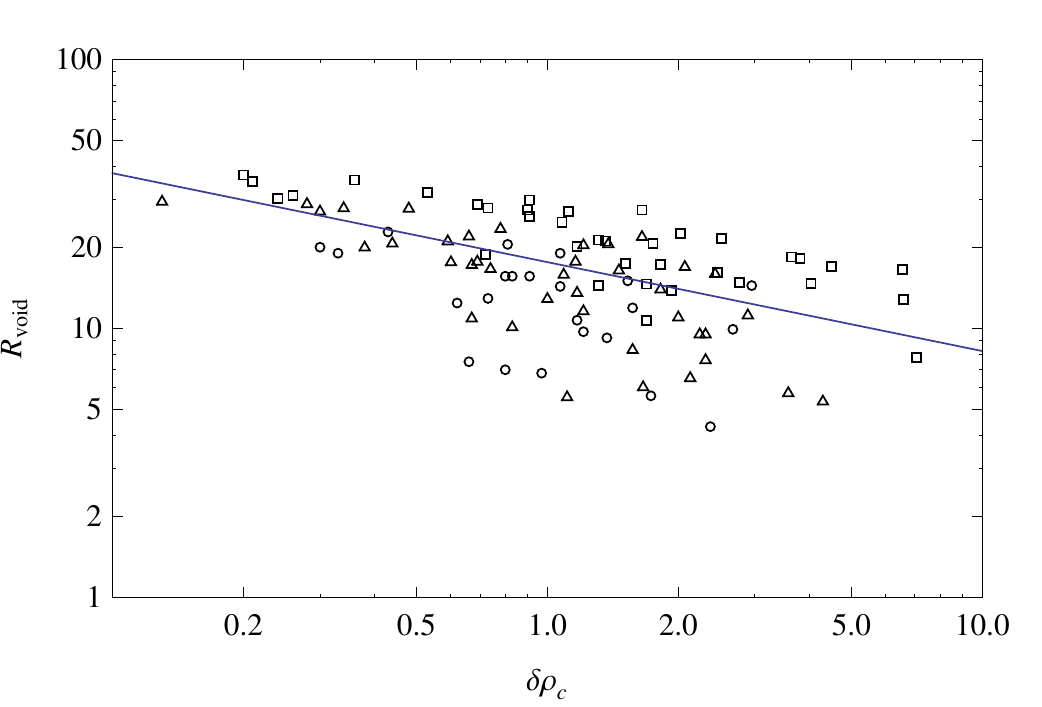}
\caption{
\label{Dvoid-rho}
Measured radii and central densities of the voids (Fig.~8 and Tables 2a-c of 
Ref.~\citet{1995A&A...301..329L}
)
 fitted to our model equation \eqref{Rvoid_vs_rhoc}. Fit is consistent with $n=3/5$ i.e. $\gamma = 8/3$.
}
\end{figure}

According to our model, radius of the void can be approximated (see Fig.~\ref{Dvoid-rho}) using:
$$
R_{void} = \frac{R_0}{\delta \rho_c^r} \quad [h^{-1} Mpc]
$$ 
with:
$$
R_0 = 17.6 \pm 0.6, \quad r=0.33 \pm 0.03 \simeq \frac{1}{3}.
$$

Polytropic exponent as large as $\gamma=8/3$ suggest that strongly
interacting matter (particles?) fill the voids. Ideal gas ($\gamma<5/3$) is excluded
in this model, because void radius would increase with central density.

\section{Conclusions }  \label{sect:Conclusions}

We have presented analysis of the spherically symmetric solutions of the TOV-$\Lambda$
equation \eqref{TOV-lambda-sys}. We focus on solution branch representing underdensity regions
in the static Einstein universe. These solutions exist only if initial
value obey $\rho_0+3 p_0<\rho_E+3 p_E$, where index $E$ refer to
values for static 3-sphere solution. 
It is shown that underdensity void-like  {\em static} regions
(in the presence of non-zero $\Lambda$ and non-trivial EOS) are possible
in general relativity. It contradicts common among cosmologist belief,
that voids can exist only because of the expansion. In principle, they could exist 
even in the static universe as well.

Static solution with density increasing outwards is the most striking
manifestation of the repulsive gravitational interaction in the presence
of the cosmological constant. These solutions remain notoriously unnoticed
and ignored, to my knowledge. For example, authors of the paper \cite{2008PhRvD..77f4008B}
seem excited with existence of the static generalization of the Einstein solution.
But in their figures 4-6 pressure (solid lines) always decrease. This is not true
for all of the solutions, cf. Fig.~\ref{TOV-lambda-examples}. We also point out that
for closed static universe there is no such thing as ''outward''(''inward'') direction.
In general, however, solutions of \eqref{TOV-lambda-dimensionless} might not be regular
for all values of $n,\alpha$ and $\lambda$. Singularity, if present, break this ''two center''
symmetry, and can be used to define ''outward direction''.

It is shown that hole repels test particles despite $\rho+3 p$ is positive inside. Instead,
$\rho+3 p - \rho_E - 3 p_E$ takes the role of gravity source,
at least in this particular model. It can be either positive or negative. Therefore
voids in static universe grow not only due to attraction of matter by structures outside
or by expansion. They can increase because test particles are expelled from  inside. This, in turn, accelerate
structure formation outside the void via ''snow plough'' mechanism of \cite{1997icm..book.....K}. 
This behavior cannot
be correctly approximated using standard N-body simulation. However, we can possibly mimic repulsive
forces with small admixture of negative mass quasi-particles in N-body codes.
These ''particles'' could represent underdensity ''holes''
(lack of the matter) similarly to holes in semiconductor physics, representing lack of the electron.
Structure formation in the static Einstein universe is of purely academic interest. Similar
effects in the real expanding universe are not excluded in the non-linear regime, however.

The static model has been applied directly to the observed profiles of voids. 
Despite the real universe is expanding, model works surprisingly well, cf. Fig.~\ref{fits}. 
Density profile can be fitted even with the simple polytropic equation of state in the framework
of the Tooper-$\Lambda$ equation \eqref{TOV-lambda-dimensionless}.  This might be
interpreted twofold: (1) physically motivated toy-model which
 works even if the real object do not satisfy all assumptions (2) manifestation
of the unknown properties of the dark matter or dark energy EOS filling the void and making
it static.

To explore the latter possibility, we have derived simplified 
version of the eq.~\eqref{TOV-lambda-dimensionless}. This ''void equation''
\eqref{void_equation} is well-suited for studying largest cosmological 
structures. It comes from a limit $\alpha \to 0$, $\lambda \to 0$ and
$\lambda/\alpha=const$ in \eqref{TOV-lambda-dimensionless} that is neither Newtonian nor
$\Lambda$-dominated. Basically, \eqref{void_equation} is the Lane-Emden 
modified by the presence of the constant $q^n$ on the right-hand-side.
The constant $q \neq 0$ break scaling symmetry present in the Lane-Emden equation.
Unexpectedly, solution for $q>0$ is far less complicated than that of Lane-Emden
equation. Asymptotic behavior (cf. \eqref{void_eq_asymptotic}) is the same for all $n$'s. 
Approximate analytical form of the solution \eqref{void_eq_approx} allow us to derive
$D_{void} - \rho_c$ relation \eqref{Rvoid_vs_rhoc}. Comparing with
data from void catalogue  \cite{1995A&A...301..329L} we get $\gamma \simeq 8/3$ in \eqref{polytropic_fluid}.

We have explored hypothesis, that the voids, are not completely empty.
 We have assumed that they are filled with unknown form of the ''dark''
matter. For simplicity, and due to poor knowledge of
 the dark matter/energy physics, polytropic EOS \eqref{polytropic_fluid}
has been used. Static model requires also that ''voids'' are already
decoupled from Hubble flow, like e.g. galaxy clusters. By extrapolation
of the \citet{2004ApJ...607..751H} result for radial density profile
to $r=0$, and from $D_{void}- \rho_c$ relation \cite{1995A&A...301..329L}
we get polytropic index in the range $0.6 \lesssim n \lesssim 1.25$.
This suggest non-ideal gas, possibly strongly interacting.
On the contrary, if we use central density as a free parameter
we obtain (i) nearly empty central region and (ii) much larger
polytropic index $1 \lesssim n \lesssim 9$. This do
not exclude ideal gas, but its nature, e.g. number of the degrees of freedom
cannot be determined.

\bibliographystyle{apsrev}
\bibliography{Hole}

\begin{thebibliography}{}

\bibitem[\protect\citeauthoryear{{B{\"o}hmer} \& {Fodor}}{{B{\"o}hmer} \&
  {Fodor}}{2008}]{2008PhRvD..77f4008B}
{B{\"o}hmer} C.~G.,  {Fodor} G.,  2008, \prd, 77, 064008

\bibitem[\protect\citeauthoryear{B\"{o}hmer, Hollenstein \& Lobo}{B\"{o}hmer
  et~al.}{2007}]{bohmer:084005}
B\"{o}hmer C.~G.,  Hollenstein L.,    Lobo F. S.~N.,  2007, Physical Review D
  (Particles, Fields, Gravitation, and Cosmology), 76, 084005

\bibitem[\protect\citeauthoryear{{B{\"o}hmer} \& {Lobo}}{{B{\"o}hmer} \&
  {Lobo}}{2009}]{2009PhRvD..79f7504B}
{B{\"o}hmer} C.~G.,  {Lobo} F.~S.~N.,  2009, \prd, 79, 067504

\bibitem[\protect\citeauthoryear{Boonserm, Visser \& Weinfurtner}{Boonserm
  et~al.}{2007}]{boonserm:044024}
Boonserm P.,  Visser M.,    Weinfurtner S.,  2007, Physical Review D
  (Particles, Fields, Gravitation, and Cosmology), 76, 044024

\bibitem[\protect\citeauthoryear{{Carneiro} \& {Tavakol}}{{Carneiro} \&
  {Tavakol}}{2009}]{2009PhRvD..80d3528C}
{Carneiro} S.,  {Tavakol} R.,  2009, \prd, 80, 043528

\bibitem[\protect\citeauthoryear{Carroll}{Carroll}{2001}]{lrr-2001-1}
Carroll S.~M.,  2001, Living Reviews in Relativity, 4

\bibitem[\protect\citeauthoryear{Ebert, Tyukov \& Zhukovsky}{Ebert
  et~al.}{2007}]{ebert:064029}
Ebert D.,  Tyukov A.~V.,    Zhukovsky V.~C.,  2007, Physical Review D
  (Particles, Fields, Gravitation, and Cosmology), 76, 064029

\bibitem[\protect\citeauthoryear{{Eddington}}{{Eddington}}{1930}]{1930MNRAS..9%
0..668E}
{Eddington} A.~S.,  1930, \mnras, 90, 668

\bibitem[\protect\citeauthoryear{Goswami, Goheer \& Dunsby}{Goswami
  et~al.}{2008}]{goswami:044011}
Goswami R.,  Goheer N.,    Dunsby P. K.~S.,  2008, Physical Review D
  (Particles, Fields, Gravitation, and Cosmology), 78, 044011

\bibitem[\protect\citeauthoryear{{Grenon}, {Elahi} \& {Lake}}{{Grenon}
  et~al.}{2008}]{2008PhRvD..78d4028G}
{Grenon} C.,  {Elahi} P.~J.,    {Lake} K.,  2008, \prd, 78, 044028

\bibitem[\protect\citeauthoryear{{Harrison}}{{Harrison}}{1967}]{1967MNRAS.137.%
..69H}
{Harrison} E.~R.,  1967, \mnras, 137, 69

\bibitem[\protect\citeauthoryear{Herrera, Ospino \& Prisco}{Herrera
  et~al.}{2008}]{herrera:027502}
Herrera L.,  Ospino J.,    Prisco A.~D.,  2008, Physical Review D (Particles,
  Fields, Gravitation, and Cosmology), 77, 027502

\bibitem[\protect\citeauthoryear{{Hled{\'{\i}}k}, {Stuchl{\'{\i}}k} \&
  {Mr{\'a}zov{\'a}}}{{Hled{\'{\i}}k} et~al.}{2004}]{2004ragt.meet...75H}
{Hled{\'{\i}}k} S.,  {Stuchl{\'{\i}}k} Z.,    {Mr{\'a}zov{\'a}} K.,  2004, in
  {Hled{\'{\i}}k} S.,  {Stuchl{\'{\i}}k} Z.,  eds, RAGtime 4/5: Workshops on
  black holes and neutron stars {Comparison of general relativistic polytropic
  and adiabatic fluid spheres with a repulsive cosmological constant}.
pp 75--89

\bibitem[\protect\citeauthoryear{{Hoyle} \& {Vogeley}}{{Hoyle} \&
  {Vogeley}}{2004}]{2004ApJ...607..751H}
{Hoyle} F.,  {Vogeley} M.~S.,  2004, \apj, 607, 751

\bibitem[\protect\citeauthoryear{{Kirshner}, {Oemler} Jr., {Schechter} \&
  {Shectman}}{{Kirshner} et~al.}{1981}]{1981ApJ...248L..57K}
{Kirshner} R.~P.,  {Oemler} Jr. A.,  {Schechter} P.~L.,    {Shectman} S.~A.,
  1981, \apjl, 248, L57

\bibitem[\protect\citeauthoryear{{Kirshner}, {Oemler}, {Schechter} \&
  {Shectman}}{{Kirshner} et~al.}{1987}]{1987ApJ...314..493K}
{Kirshner} R.~P.,  {Oemler} A.~J.,  {Schechter} P.~L.,    {Shectman} S.~A.,
  1987, \apj, 314, 493

\bibitem[\protect\citeauthoryear{{Krasinski}}{{Krasinski}}{1997}]{1997icm..boo%
k.....K}
{Krasinski} A.,  1997, {Inhomogeneous Cosmological Models}.
Cambridge University Press

\bibitem[\protect\citeauthoryear{Lake}{Lake}{2003}]{PhysRevD.67.104015}
Lake K.,  2003, Phys. Rev. D, 67, 104015

\bibitem[\protect\citeauthoryear{{Lake}}{{Lake}}{2008}]{2008PhRvD..77l7502L}
{Lake} K.,  2008, \prd, 77, 127502

\bibitem[\protect\citeauthoryear{{Lake}}{{Lake}}{2009}]{2009arXiv0905.3546L}
{Lake} K.,  2009, ArXiv e-prints

\bibitem[\protect\citeauthoryear{{Lindner}, {Einasto}, {Einasto}, {Freudling},
  {Fricke} \& {Tago}}{{Lindner} et~al.}{1995}]{1995A&A...301..329L}
{Lindner} U.,  {Einasto} J.,  {Einasto} M.,  {Freudling} W.,  {Fricke} K.,
  {Tago} E.,  1995, \aap, 301, 329

\bibitem[\protect\citeauthoryear{Martin \& Visser}{Martin \&
  Visser}{2004}]{PhysRevD.69.104028}
Martin D.,  Visser M.,  2004, Phys. Rev. D, 69, 104028

\bibitem[\protect\citeauthoryear{{Parisi}, {Bruni}, {Maartens} \&
  {Vandersloot}}{{Parisi} et~al.}{2007}]{2007CQGra..24.6243P}
{Parisi} L.,  {Bruni} M.,  {Maartens} R.,    {Vandersloot} K.,  2007, Classical
  and Quantum Gravity, 24, 6243

\bibitem[\protect\citeauthoryear{{Peebles}}{{Peebles}}{2001}]{2001ApJ...557..4%
95P}
{Peebles} P.~J.~E.,  2001, \apj, 557, 495

\bibitem[\protect\citeauthoryear{Robertson}{Robertson}{1933}]{RevModPhys.5.62}
Robertson H.~P.,  1933, Rev. Mod. Phys., 5, 62

\bibitem[\protect\citeauthoryear{{Seahra} \& {B{\"o}hmer}}{{Seahra} \&
  {B{\"o}hmer}}{2009}]{2009PhRvD..79f4009S}
{Seahra} S.~S.,  {B{\"o}hmer} C.~G.,  2009, \prd, 79, 064009

\bibitem[\protect\citeauthoryear{{Stuchlik}}{{Stuchlik}}{2000}]{Stuchlik}
{Stuchlik} Z.,  2000, Acta Physica Slovaca, 50, 219

\bibitem[\protect\citeauthoryear{Tegmark, Strauss, Blanton, Abazajian,
  Dodelson, Sandvik, Wang, Weinberg, Zehavi, Bahcall \& Hoyle}{Tegmark
  et~al.}{2004}]{PhysRevD.69.103501}
Tegmark M.,  Strauss M.~A.,  Blanton M.~R.,  Abazajian K.,  Dodelson S.,
  Sandvik H.,  Wang X.,  Weinberg D.~H.,  Zehavi I.,  Bahcall N.~A.,    Hoyle
  F.,  2004, Phys. Rev. D, 69, 103501

\bibitem[\protect\citeauthoryear{{Tooper}}{{Tooper}}{1965}]{1965ApJ...142.1541%
T}
{Tooper} R.~F.,  1965, \apj, 142, 1541

\bibitem[\protect\citeauthoryear{{Wu} \& {Yu}}{{Wu} \&
  {Yu}}{2009}]{2009arXiv0909.2821W}
{Wu} P.,  {Yu} H.,  2009, ArXiv e-prints

\bibitem[\protect\citeauthoryear{{Zeldovich}, {Einasto} \&
  {Shandarin}}{{Zeldovich} et~al.}{1982}]{1982Natur.300..407Z}
{Zeldovich} I.~B.,  {Einasto} J.,    {Shandarin} S.~F.,  1982, \nat, 300, 407

\end{thebibliography}

\end{document}